\documentclass[prd,twocolumn,showpacs,preprintnumbers,amsmath,amssymb]{revtex4}
\usepackage{mathrsfs}
\usepackage{graphicx}
\usepackage{epsfig}
\usepackage{dcolumn}
\usepackage{bm}

\let\jnfont=\rm

\def\NPB#1,{{\jnfont  Nucl.\ Phys.\ B }{\bf #1},}
\def\PLB#1,{{\jnfont Phys.\ Lett.\ B }{\bf #1},}
\def\EPJC#1,{{\jnfont Euro.\ Phys.\ J.\ C }{\bf #1},}
\def\PRD#1,{{\jnfont \em Phys.\ Rev.\ D }{\bf #1},}
\def\PRL#1,{{\jnfont Phys.\ Rev.\ Lett.\ }{\bf #1},}
\def\MPLA#1,{{\jnfont Mod.\ Phys.\ Lett.\ A }{\bf #1},}
\def\JPG#1,{{\jnfont J.\ Phys.\ G}{\bf #1},}
\def\CTP#1,{{\jnfont Commun.\ Theor.\ Phys.\ }{\bf #1},}
\def\CPC#1,{{\jnfont Chin. \ Phys. \ C }{\bf #1},}

\def\p_slash{\not{\hbox{\kern-2.1pt $p$}}}
\def\k_slash{\not{\hbox{\kern-2.1pt $k$}}}
\def\overlay#1#2{\ifmmode \setbox 0=\hbox {$#1$}\setbox 1=\hbox to\wd 0{\hss
$#2$\hss }\else \setbox 0=\hbox {#1}\setbox 1=\hbox to\wd 0{\hss
#2\hss }\fi #1\hskip -\wd 0\box 1}

\begin{document}

\title{Top quark decay $t \to cb\bar{b}$ in topcolor-assisted technicolor models }

\author{Guo-Li Liu}

\affiliation{
Department of Physics, Zhengzhou University, ZhengZhou, Henan 450001, China; \\
Kavli Institute for Theoretical Physics China, Academia Sinica, Beijing 100190, China
     \vspace*{0.5cm}}

\begin{abstract}
Topcolor-assisted technicolor (TC2) model predicts the existence of
the top-pions and the CP-even top-Higgs with large flavor-changing
couplings to the top quark, which at tree-level can mediate the top
quark three-body decay $t \to cb\bar{b}$.
In this work we study this decay, showing the dependence
of the decay rate on the relevant TC2 parameters and comparing the
results with the predictions in the minimal supersymmetric model.
We find that the decay rate in the TC2 model is much larger than
in the minimal supersymmetric model and, in a large part of the
parameter space, the TC2 prediction may reach the detectable level
of the LHC.
\end {abstract}


\maketitle

\section{Introduction}
The top quark was discovered at the Tevatron collider, but its properties
have not been precisely measured due to the small statistics of this collider.
The large hadron collider (LHC) at CERN will operate as a top quark
factory, producing about eight millions of top-pair events per
year in its first stage, and thus will scrutinize
the top quark properties \cite{TOPReviews}. Any new physics related to top
quark will be uncovered or stringently constrained \cite{sensitive}.

In the SM the extraordinary large mass of the top quark renders the GIM
mechanism very effective in the top quark sector even at loop level.
As a result, the top quark flavor-changing neutral-current (FCNC)
interactions are extremely small in the SM \cite{tcvh-sm}, which
implies that the observation of any FCNC top quark process could be
a robust evidence for new physics beyond the SM. This has motivated
intensive studies for the  FCNC top quark interactions in various
new physics models,  such as the popular minimal supersymmetric
model (MSSM) \cite{tcv-mssm}
and the topcolor-assisted technicolor (TC2) model \cite{tcv-tc2}.
The studies in the literature are so far concentrating on the
loop-induced FCNC processes of the top quark. In this work
we focus on the three-body decay $t\to cb\bar{b}$, which can be
mediated by a charged Higgs boson in the MSSM or
by a top-pion/top-Higgs in the TC2 model.

The TC2 model predicts the existence of the top-pions ($\pi _{t}^{\pm},\pi _{t}^{0}$)
and the top-Higgs $ h_{t}^{0}$ at the weak scale \cite{tc2}. These scalars
have large flavor-changing couplings to the top quark:
besides the large charge-current coupling of $\pi _{t}^{\pm} t \bar b$,
the FCNC couplings $\pi _{t}^{0} t\bar c$ and $h_{t}^{0} t\bar c$
occur at tree-level and can also be large.
These flavor-changing couplings can induce the three-body decay
$t\to cb\bar{b}$ mediated by a top-pion or top-Higgs at tree level.
In our study we will show the dependence
of the decay rate on the relevant TC2 parameters and compare the
results with the predictions in the MSSM.

This work is organized as follows. We will briefly discuss the TC2 model
in Section II, giving the new couplings which will be involved in our
calculation. In Section III we give the calculation results and compare
with the result in the MSSM and the SM.
Finally, the conclusion is given in Section IV.

\section{About TC2 model}
To solve the phenomenological difficulties of the traditional TC theory,
TC2 theory \cite{tc2} was proposed by combining technicolor (TC)
interactions with the topcolor interactions at the TeV scale.
In TC2 theory, the TC interactions play a main role in breaking the
electroweak symmetry. The ETC interactions give rise to the masses
of the ordinary fermions including a very small portion of the top
quark mass, namely $\epsilon m_{t}$ with a model dependent parameter
$\epsilon \ll 1$. The topcolor interactions also make small
contributions to the EWSB, but its main role is to give rise to the main
part of the top quark mass $(1-\epsilon)m_{t}$.

At the weak scale the TC2 model predicts the existence of two groups of
scalars from topcolor and technicolor condensations \cite{tc2,top-condensation}.
In the linear realization, the scalars can be arranged into two $SU(2)$
doublets, namely  $\Phi_{top}$ and $\Phi_{TC}$
\cite{top-condensation,2hd,Rainwater}, which are analogous to the
Higgs fields in the two-Higgs-doublet model. The
doublet $\Phi_{top}$ from topcolor condensation couples only to the
third-generation quarks, whose main task is to generate the large
top quark mass. It can also generate a sound part of bottom quark
mass indirectly via instanton effect \cite{tc2}. Since a small value
of the top-pion decay constant $F_t $ (the VEV of the doublet
$\Phi_{top}$) is theoretically favored, this doublet must couple
strongly to top quark in order to generate the expected top quark
mass. The other doublet $\Phi_{TC}$, which is technicolor
condensate, is mainly responsible for EWSB and the generation of
light fermion masses. It also contributes a small portion to the
third-generation quark masses.  However, its Yukawa couplings
with all fermions are small because its VEV $v_{TC}$ is generally
comparable with $v_W$. The flavor changing Yukawa couplings of the
new scalars $\pi_t^\pm$, $\pi_t^0$ and $h_t^0$ are given by \cite{tc2}
\begin{eqnarray}
\cal{L}&=&\frac{(1 - \epsilon ) m_{t}}{\sqrt{2}F_{t}}
       \frac{\sqrt{v_{w}^{2}-F_{t}^{2}}} {v_{w}}  \left(
            \sqrt{2}K_{UR}^{tt *} K_{DL}^{bb}\bar{t}_R b_{L}\pi_t^+ \right. \nonumber \\
    && + \sqrt{2} K_{UR}^{tc *} K_{DL}^{bb} \bar{c}_R b_{L} \pi_t^+
           + iK_{UR}^{tc}K_{UL}^{tt*}\bar{t}_L c_{R}\pi_{t}^{0}
            \nonumber \\
    && \left.+ K_{UL}^{tt *} K_{UR}^{tc} \bar{t}_L c_{R} h_t^0 + h.c.  \right) ,
\label{FCNH}
\end{eqnarray}
where we neglected the mixing between up quark and top quark, and
$K_{UL}$, $K_{DL}$ and $K_{UR}$ are the rotation matrices that
transform the weak eigenstates of left-handed up-type, down-type and
right-handed up-type quarks to their mass eigenstates, respectively.
Their favored values are given by \cite{tc2}
\begin{eqnarray}
&K_{UL}^{tt}& \simeq K_{DL}^{bb} \simeq 1, \hspace{5mm}
 K_{UR}^{tt}\simeq \frac{m_t^\prime}{m_t} = 1-\epsilon, \nonumber \\
&K_{UR}^{tc}&\leq \sqrt{1-(K_{UR}^{tt})^2} =\sqrt{2\epsilon-\epsilon^{2}},
\label{FCSI}
\end{eqnarray}
with $m_t^\prime$ denoting the topcolor contribution to the top
quark mass.

The couplings of top-pion and top-Higgs with $b\bar{b}$ are given by
\begin{eqnarray}
   \cal{L'}&&=  \frac{m_b^*}{\sqrt{2}F_{t}}\frac{\sqrt{v_{w}^{2}-F_{t}^{2}}} {v_{w}}(i\bar{b}
   b\pi_{t}^{0}+\bar{b}b h_{t}^{0}),
\label{NFC}
\end{eqnarray}
where $m_b^* (\leq m_b)$ is the bottom quark mass created by instatons
and is approximately given by
\begin{equation}
 m^*_b \, \approx \, {3 k m_t \over 8 \pi^2} \simeq \, 6.6 \, k \; {\rm GeV} \; .
 \label{eq:mb}
\end{equation}
To get a limit on $k$, we use a bottom quark pole mass of $m_b
\approx 5$ GeV, so that the entire bottom quark mass would come
from contribution of topcolor instantons for $k \sim 0.73$.
Here we use $k = 0.61$ for $m_b=5$ GeV.
The remaining $m_b$ contribution is assumed to come from ETC via a
Yukawa coupling $\epsilon_b$.

\section{Top decay $t\to c\bar{b}b$ in TC2 model}
\begin{figure}[hbt]
\begin{center}
\epsfig{file=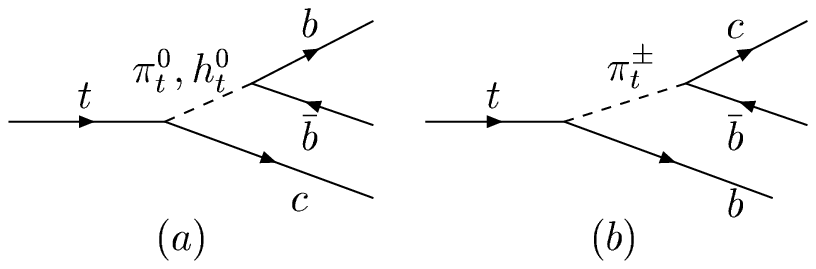,width=8cm}
\caption{ Feynman diagrams of $t\to c\bar{b}b$ in TC2 model. }
\label{feynman}
\end{center}
\end{figure}

The Feynman diagrams of the process $t\to c\bar{b}b$
 mediated by a top-pion or top-Higgs at tree level are given in Fig.~\ref{feynman}.
The relevant couplings can be found in Eqs.(\ref{FCNH}) and (\ref{NFC}).
The amplitude of this process is given by
\begin{eqnarray}
{\cal M}&=&\frac{m_t}{2F_t}\frac{\sqrt{V_W^2-F_t^2}}{V_W} K_{UR}^{tt*}K_{UR}^{tc}  \nonumber \\
&&\times \left[ \frac{im_b^*}{(p_b+p_{\bar{b}})^2
         -m_{\pi^0_t}^2}\bar{u}_b\gamma_5v_b \bar{u}_cP_Lu_t \right.\nonumber \\
&&+\frac{-m_b^*}{(p_b+p_{\bar{b}})^2-m_{h^0_t}^2}\bar{u}_b\gamma_5v_b \bar{u}_cP_Lu_t\nonumber \\
&&\left.+\frac{-2m_t}{(p_b+p_{\bar{b}})^2-m_{\pi^{\pm}_t}^2}\bar{u}_b\gamma_5v_b \bar{u}_cP_Lu_t \right],
\end{eqnarray}
where $p_b$($p_{\bar{b}})$ denotes the momentum of the
$b$ ($\bar{b}$) quark in the final state and
$P_{R,L}=(1\pm\gamma^5)/2$ denotes the chiral operator.

The decay rate of $t \to cb\bar{b}$ in TC2 model
depends on the parameter $\epsilon$ (which varies in the range of $0.01-0.1$)
and the masses of the top-pions and top-Higgs.
Since the mass splitting between the neutral top-pion and the
charged top-pion comes only from the electroweak interactions and
thus should be small,  we assume
$m_{\pi_{t}^{0}}=m_{\pi_{t}^{\pm}}\equiv m_{\pi_t}$.
The top-pion masses are allowed to be a few hundred GeV \cite{pion-mass},
depending on the details of the considered models. The top-Higgs mass
can lie in the same range as the top-pion masses. If we assume
the mass degeneracy for top-Higgs and top-pions,
we can just write $M_{TC}$($=m_{\pi_t}=m_{h_t}$) to denote their common mass.
As for other involved parameters, we take $m_{t}=172$ GeV \cite{topmass},
$v_W=174$ GeV, $F_t=50$ GeV, $m_b=5$ GeV,  $m_b^*=4$ GeV and  $m_c=1$ GeV.
\begin{figure}[hbt]
\begin{center}
\epsfig{file=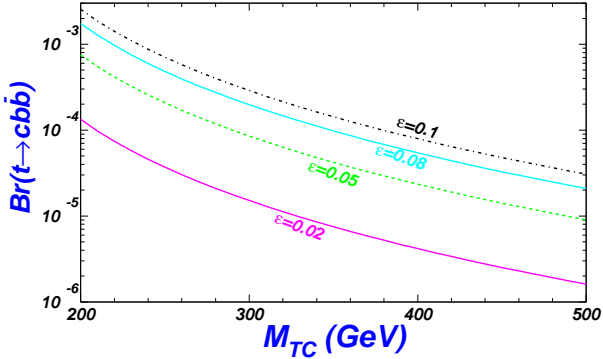,width=8cm}
\caption{The branching ratio of $t\to c\bar{b}b$ versus $M_{TC}$ in TC2 model. }
\label{mtc}
\end{center}
\end{figure}

Fig.\ref{mtc} shows the  branching ratio of $\Gamma(t\to
cb\bar{b})$ as a function of the scalar mass $M_{TC}$ for four
values of the parameter $\epsilon$. From this figure we can see
that the decay branching ratio decreases with $M_{TC}$, showing
the decoupling effects of the heavy top-pions/top-Higgs. As
$\epsilon$ increases, the decay branching ratio increases. The
reason is that the couplings $\pi^\pm_t c b$, $\pi^0_t t c$ and
$h^0_t t c$ are all proportional to $\epsilon$. Note that here we
choose  $M_{TC}>m_t$ so the scalars cannot be on shell. Actually,
so far the top-Higgs can still be possibly lighter than top quark
while the top-pions are relatively heavy \cite{tcv-tc2}:
\begin{equation}
m_{h_t^0} > 135 ~{\rm GeV}, \hspace{3mm}
m_{\pi_{t}^{0}}=m_{\pi_{t}^{\pm}}\equiv m_{\pi_t} > 220 ~{\rm GeV} .
\end{equation}
So for $m_t >m_{h_t^0}+m_c$, the top-Higgs $h_t^0$ in $t\to ch_t^0 \to cb\bar{b}$
can be on-shell. To calculate the decay rate of $h_t^0\to b \bar{b}$, we need to
compute the total decay width of the top-Higgs. Its possible decay channels are
\begin{eqnarray}
h_t^0 \to  t\bar{t}, ~t\bar{c}, ~\bar{t} c, ~b\bar{b},
       ~WW,   ~Z Z,  ~\gamma Z,  ~g g, ~\gamma \gamma.
\label{decay}
\end{eqnarray}
\begin{figure}[hbt]
\begin{center}
\epsfig{file=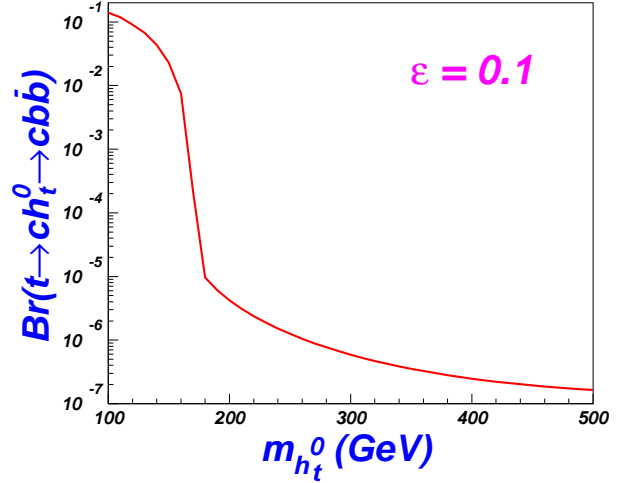,width=8cm}
\caption{The branching ratio of $t\to c h_t^0\to c\bar{b}b$ versus $m_{h_t^0}$ in TC2 model.}
\label{mh}
\end{center}
\end{figure}

In Fig.\ref{mh} we plot the contribution of $h_t^0$ to the
branching ratio of $t \to c\bar{b}b$.  As expected, when  $h_t^0$
is light enough to be on-shell, the decay branching ratio can be
very large.  The reason is that when $h_t^0$ is on-shell,
the two-body decay $t\to c h_t^0$ has a large rate and,
at the same time, the decay $h_t^0\to b\bar b$ is the diminant mode
for a light $h_t^0$.

Note that in the decay  $t \to c\bar{b}b$ the top-Higgs contribution is
dominant only for a light top-Higgs (so it can be on-shell).
When the top-Higgs and top-pions are assumed to be degenerate and
heavier than the top quark mass, the charged top-pion contribution is
dominant.

\begin{figure}[hbt]
\begin{center}
\epsfig{file=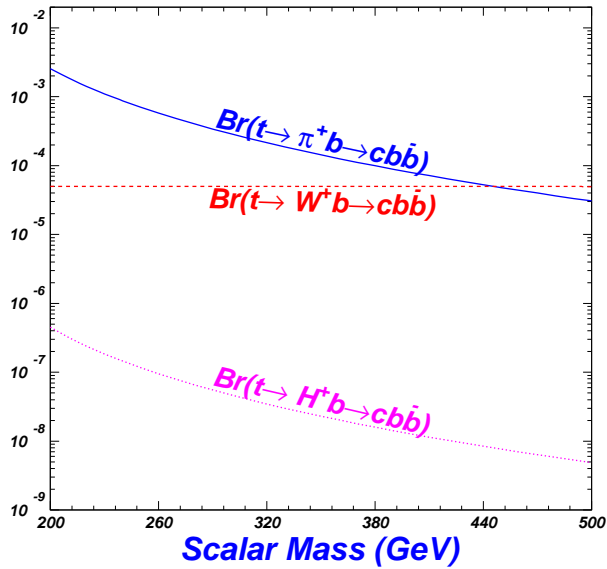,width=8cm}
\caption{ The solid curve is $Br(t \to \pi^+_t b\to c\bar{b}b)$
versus $M_{\pi^+_t}$ in TC2.  The dashed line is $Br(t \to W^+ b\to
c\bar{b}b)$ in the SM. The dotted curve is  $Br(t \to H^+ b\to
c\bar{b}b)$ versus $M_{H^+}$ in the MSSM. } \label{compare}
\end{center}
\end{figure}

In the MSSM the decay  $t \to c\bar{b}b$ can be mediated by a
charged Higgs boson. However, the coupling of the charged Higgs
boson with $c\bar{b}$ is small, $\sim
\frac{ieV_{cb}}{2\sqrt{2}m_Wsin\theta_W}m_b\tan\beta P_L$, which is
suppressed by both $m_b/m_W$ and $V_{cb}$. In the SM, $t \to
c\bar{b}b$ can be mediated by a $W$-bsoson. In Fig.\ref{compare} we
compare the three channels for the decay $t \to c\bar{b}b$. We see
that $Br(t \to \pi^+_t b\to c\bar{b}b)$ in TC2 model is much larger
than $Br(t \to H^+ b\to c\bar{b}b)$ in the MSSM, where we take
$\tan\beta=40$, which is quite large. For a light top-pion, $Br(t
\to \pi^+_t b\to c\bar{b}b)$ can be much larger than $Br(t \to W^+
b\to c\bar{b}b)$ in the SM.

Note that the decay  $t \to c\bar{b}b$  can also be mediated by a
vector boson, i.e., $t \to cV \to c \bar{b}b$ ($V=g,\gamma,Z$) with
the loop-induced vertex $tcV$.
Although such loop-induced $tcV$ couplings can be greatly enhanced in
TC2 model or the MSSM \cite{tcv-mssm,tcv-tc2}, their contribution
to $t \to c \bar{b}b$ is much smaller than the tree-level diagrams
in Fig.1.

Since the decay  $t \to c\bar{b}b$ can have a quite large branching
ratio in the TC2 model, it would be accessible at future colliders
like the LHC and the ILC. Such rare decays can be searched from the
$t\bar{t}$ events with one top decaying in the rare mode while the
other top having the SM decay $t \to W b \to \ell \nu b$ ($\ell=e,
\mu$) \cite{saavedra}. The discovery reach of the rare top quark
decays in the future colliders for $100fb^{-1}$ of integrated
luminosity is roughly given by\cite{sola}
\begin{eqnarray}
 LHC   :B_{r}(t\to cX) \geq 5\times10^{-5} \nonumber \\
  \hspace{7mm} LC    :B_{r}(t\to cX) \geq 5\times10^{-4} \nonumber \\
  \hspace{7mm}TEV33 :B_{r}(t\to cX) \geq 5\times10^{-3}
\end{eqnarray}
Therefore, the effects of top-pions and top-higgs in the rare top decay
$t\to c b\bar{b}$ might be experimentally accessible at the LHC.

\section{Conclusion}
We calculated the top quark three-body decay $t \to cb\bar{b}$ mediated
by a top-pion or top-Higgs in the TC2 model.
We showed the dependence
of the decay rate on the relevant TC2 parameters and compared the
results with the predictions in the minimal supersymmetric model.
We found that the decay rate in the TC2 models is much larger than
in the minimal supersymmetric model and, in a large part of the
parameter space, the TC2 prediction may reach the detectable level
of the LHC.

\end{document}